\documentclass[aps,prx,twocolumn,groupedaddress]{revtex4-2}
\usepackage{graphicx}
\usepackage{amsmath}
\usepackage{amssymb}
\usepackage{braket}
\usepackage{hyperref}
\usepackage{xcolor}
\usepackage{enumitem}  
\hypersetup{
	colorlinks,
	linkcolor={blue},
	citecolor={blue},
	urlcolor={blue}
}
\newcommand{\pardev}[2]{\dfrac{\partial{#1}}{\partial{#2}}}

\newcommand{\simpleSum}{\sum_{k=1}^N}

\newcommand{\sethighlightcolor}[1]{%
	\ifnum#1=0 
	\definecolor{highlightcolor}{RGB}{0,0,0}%
	\else 
	\definecolor{highlightcolor}{RGB}{0,130,0}%
	\fi
}

\newcommand{\highlight}[1]{%
	\textcolor{highlightcolor}{#1}%
}

\sethighlightcolor{0}

\begin{document}
\title{Inverse design of arbitrary optical helicity patterns}
\author{Romuald Kilianski}
\email[email: ]{r.kilianski.1@research.gla.ac.uk}
\author{Robert Bennett}
\affiliation{School of Physics \& Astronomy, University of Glasgow, Glasgow G12 8QQ, United Kingdom}

\date{\today}

\begin{abstract}
Superposing multiple plane waves can generate helicity lattices in which the optical helicity varies regularly in space. Here we propose an inverse design method for constructing arbitrary helicity structures based on placing a digital object consisting of dielectric inclusions in three-dimensional space. We apply the method to design structures that reproduce two-dimensional lattices embedded within a three-dimensional region using only a single plane wave as an input. In order to demonstrate the power and flexibility of our method, we go beyond the paradigm of a regular lattice and propose structures which can create arbitrary images consisting of regions of varying helicity, again using only a single plane wave as an input. 
\end{abstract}

\maketitle

\section{Introduction}
\highlight{The concept of chirality appears across the sciences and details the characteristics of geometrical arrangements with identical constituents but inverted, non-superimposable forms. Circularly polarised light fits the criteria of a chiral object as spatial inversion transforms left-handed circularly polarised light into its right-handed form --- the chiral structure is the combination of its propagation direction and polarisation. This property makes light with circular polarisation exert influence on other chiral objects with which it interacts. For example, the dependence of a chiral molecule's absorption cross section on the type of circularly polarised light that illuminates it is described by an effect known as circular dichroism (see, for example, \cite{barron2009molecular}). In essence, the rate of excitation is sensitive to the molecule's handedness and that of the illuminating light. The degree to which an electromagnetic field influences such an interaction can be related to a product of the field vectors, a parity-odd scalar density called optical chirality \cite{tang_cohen} (originally introduced by Lipkin as the scalar component of the `00-zilch' \cite{lipkinEXISTENCENEWCONSERVATION1964}, but not assigned a physical interpretation at that early stage). In our example, we are interested in monochromatic fields. For this case, the optical chirality is equal --- up to a constant \cite{Mackinnon_2019} --- to a quantity called \emph{electromagnetic helicity} \cite{truebaElectromagneticHelicity1996}. Familiar from particle physics, as well as analogous quantities in plasma physics \cite{plasma}, helicity describes the projection of the spin angular momentum onto the direction of propagation. Our specific focus is the electromagnetic helicity \emph{density} (see, e.g., $\mathcal{H}$ \cite{Cameron_2012}), which for a monochromatic plane wave of frequency $\omega$ reads
\begin{align}\label{helicity}
\mathcal{H} = 	-\dfrac{i}{4 c \omega}  \left( \mathbf{E} \cdot \mathbf{H^{*}} - \mathbf{E^{*}} \!\! \cdot \mathbf{H}\right),
\end{align}
where $\mathbf{E}$ and $\mathbf{H}$ are the complex electric and magnetic fields, respectively, and $c$ is the speed of light. }

\subsection{Helicity lattices}

\highlight{We will consider circumstances where the helicity density given by Eq.~\eqref{helicity} varies regularly in space and forms a so-called helicity lattice \cite{jorg}, a periodic structure resembling the arrangement of intensity patterns in optical lattices. Helicity lattices are defined by having a helicity density that varies in space but an electric field intensity that remains constant --- the latter feature is of particular importance when considering the interactions of chiral matter with the lattice. A chiral optical force mediated by the helicity gradients in the field can accelerate enantiomers (molecular species of opposite handedness) in different directions \cite{cameron2017chirality}, providing a means of separating them. Importantly, the homogeneity of the electric field intensity in a helicity lattice means that achiral optical forces are (in principle) completely suppressed, allowing the much weaker chirality-dependent forces to be isolated. Oppositely handed species often exhibit drastically different properties, so that methods for efficiently separating and distinguishing them are an active area of research --- for a comprehensive review, see \cite{Separating_en_review}. Helicity lattices are already finding their place in this field, for example a scheme was recently proposed to utilise helicity lattices in separating cold, chiral molecules via their chirality-sensitive quantum phase transitions \cite{Isaule_2022}. } 

\highlight{With their potential in industrial applications, the study of helicity lattices is ready to move from theoretical curiosity towards an exploration of more practical methods for their generation. In their original formulation \cite{jorg}, helicity lattices were discovered by superimposing up to six idealised plane waves with particular well-defined polarisations. Aside from the impossibility of generating a true plane wave in the laboratory, the need to align and manipulate up to six differently-polarised beams is a significant technical challenge --- our goal in this work is to define a proof-of-principle method for generating helicity lattices in a more robust way. The problem is similar to that encountered in generating magneto-optical trapping potentials, where again up to six beams can be required. A way of circumventing this issue in that context has been to use reflective or diffractive optical elements to generate the required number of beams in an integrated and compact apparatus (see, for example, \cite{nshiiSurfacepatternedChipStrong2013, barkerSingleBeamZeemanSlower2019, zhuDielectricMetasurfaceOptical2020}). While it may indeed be possible to follow similar design principles to create a helicity lattice, the additional complexity found therein (especially regarding polarisation-sensitivity) leads us to take a different approach, employing the techniques of \emph{inverse design}.  }

\subsection{Inverse design}

\highlight{Traditional `forward' design of optical components relies on a human designer using some combination of general principles, intuition and previous experience to produce a candidate design which is then tested and refined against some desired figure of merit. Inverse design turns this paradigm around --- only the figure of merit is specified, following which a computer algorithmically builds up an appropriate device. The vast space of possible designs originally necessitated some imposition of geometrical constraints in early work in this direction \cite{dobson1999maximizing,earlyfelici2001shape}, for example the sizes and positions of fixed circles and rectangles would be altered such that an improved design is reached. The introduction to photonics of \emph{adjoint methods} from aerospace engineering \cite{Volpe1986TheDO}, coupled with increased computing power, allowed such constraints to be lifted, kicking off the era of completely free-form photonic inverse design \cite{borel2004topology,jensen2004systematic}. The combination of this and the relatively recent ability to manipulate matter on the micro- and nano-scales has initiated an explosion in the application of inverse design methods (for a comprehensive review, see \cite{Molesky}). In contrast to the earlier constrained approaches, the pool of available designs is unlimited --- the shapes of the structures are dictated by their impact on performance (and manufacturability) and are no longer guided by intuition or arbitrary constraints. Some of the more recent diverse applications of these ideas have been concerned with the conversion of solar energy \cite{solar}, integrated photonic devices \cite{OnChip}, sub-wavelength focusing \cite{dobson2009optimization} and heat transfer \cite{thermal}. Most recently, ID has been applied to atomic coherence \cite{RobAtomicTran}, light-matter interactions \cite{RobShape} and nuclear quantum optics \cite{XRAYID}.}

\highlight{As discussed in the previous sections,} it has been shown analytically in Ref.~\cite{jorg} that helicity lattices can be constructed from up to six idealised plane waves in various configurations, \highlight{and that we would like to come up with a way of relaxing this requirement.} To accomplish this, we \highlight{will use a freeform inverse design approach to generate} a dielectric structure that can refract a single beam multiple times in such a way that a lattice structure emerges in a given plane. The placement of the elements forming such a structure \highlight{will turn out to be} non-obvious and \highlight{not easily} inferred from a wave analysis point of view. The desired outcome will be chosen, and an algorithm will adjust the geometry to \highlight{bring a chosen merit function as close as possible to} a designated goal. 

This paper is organised as follows: in Section \ref{DescriptionOfProblemSection}, we will set up the problem of helicity lattice generation in terms of an optimisation task, develop the mathematical structure and propose an algorithm capable of reproducing a helicity lattice. In Section \ref{ProcedureSection} we will then outline the specifics of the actual inverse design process. Subsequently, in Section \ref{ResultsSection} we will present the results of two single-beam simulations reproducing lattices originally formed by a superposition of three sources. Additionally, in that section we present a novel approach where we simulate a helicity density that is not periodic and can take an arbitrary (user-defined) shape. \highlight{Critical analysis of the method and conclusions follow in sections and \ref{analysisSection} and \ref{conclusionsSection}.}

\section{Description of the problem}\label{DescriptionOfProblemSection}
\begin{figure}
	\centering
	\includegraphics[width=1\linewidth]{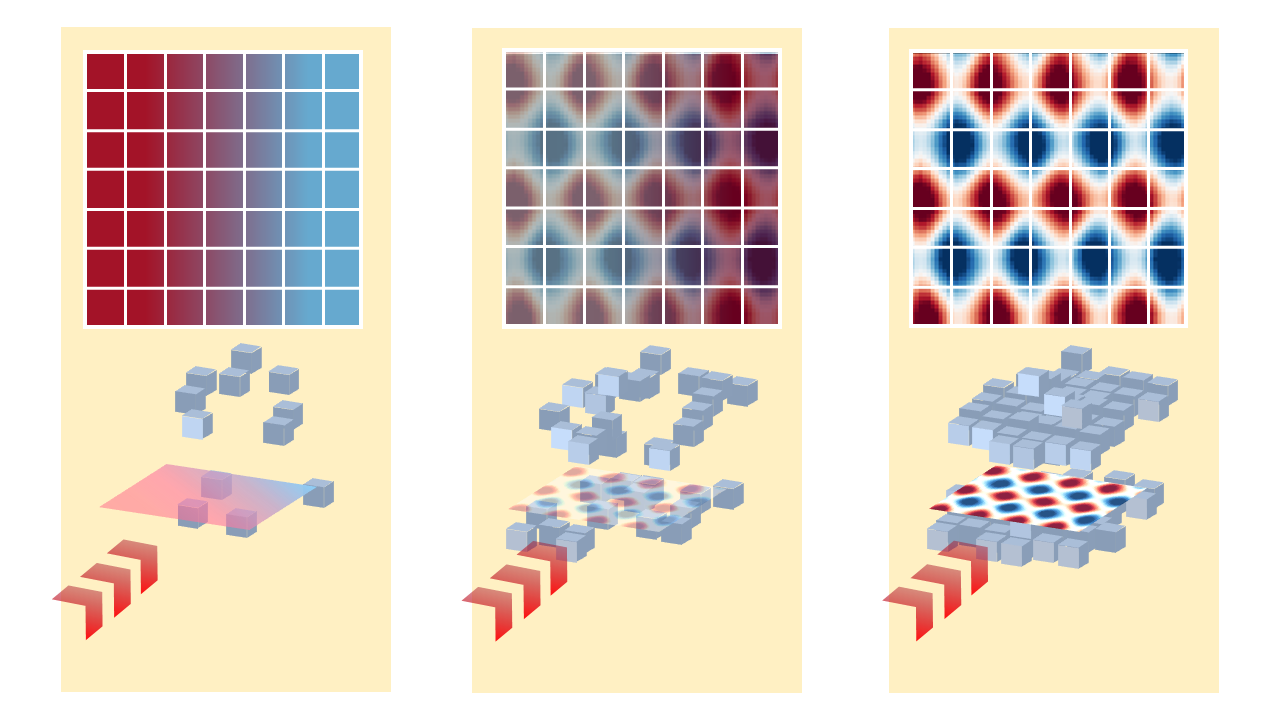}
	\caption{The schematic of different stages of building the dielectric structure using a single beam source. As it grows in size and complexity, the more the underlying field approximates the helicity lattice. }
	\label{fig:helicity3stages}
\end{figure}
We aim to produce a two-dimensional optical pattern resembling a helicity lattice \highlight{as closely as possible}. Instead of superposing multiple light sources, we want to effect it by shining a single beam onto a crystal-like (not uniformly periodic) structure that will refract the beam in such a way to ensure that a helicity lattice emerges. Our idea is to create such a structure algorithmically; instead of iterating over possible geometries, we let an algorithm grow the shape of the crystal. \highlight{This idea is shown schematically in Fig.~\ref{fig:helicity3stages}, where we see the underlying field progressively changing and starting to resemble the target lattice as more inclusions are added.}  At every step, the algorithm will decide on the optimal placement of a dielectric inclusion --- we will use an inverse design algorithm based on the adjoint method. 

\highlight{The adjoint method relies on the fact that Maxwell's equations in (reciprocal) media are symmetrical with respect to the exchange of source and observation points; this is known as Onsager reciprocity \cite{onsagerReciprocalRelationsIrreversible1931}. This concept is most naturally expressed in terms of the (dyadic) Green's tensors that solve Maxwell's equation via the inhomogeneous Helmholtz equation.} The correspondence between the sources in terms of Green's tensors has been outlined in, for example, \cite{Miller:EECS-2012-115}, thus we are going to state the important results which form the foundational symmetry argument used in the adjoint method. The Cartesian component $ij$ of a Green's tensor describing an electric field \highlight{at $\mathbf{r}_1$ that originates} from a point-like electric dipole \highlight{source at $\mathbf{r}_2$} can be written as $\mathcal{G}^{\mathrm{EP}}_{ij}(\mathbf{r}_{1},\mathbf{r}_{2})$. Analogously, the elements of a Green's tensor relating magnetic source with its field are $\mathcal{G}^{\mathrm{HM}}_{ij}(\mathbf{r}_{1},\mathbf{r}_{2})$. We can express the source-observer symmetry in reciprocal media in terms of $\mathcal{G}^{\mathrm{EP/HM}}$ as:
\begin{equation}
\highlight{\mathcal{G}^{\mathrm{EP/HM}}_{ij}(\mathbf{r}_{2},\mathbf{r}_{1}) = \mathcal{G}^{\mathrm{EP/HM}}_{ji} (\mathbf{r}_{1},\mathbf{r}_{2}).\label{EE}}
\end{equation}
\highlight{The $i$-th component of an electric/magnetic field at $\mathbf{r}_2$ arising from the $j$-th component of an electric/magnetic dipole at $\mathbf{r}_1$ is equal to the $j$-th component of an electric/magnetic field at  $\mathbf{r}_1$ from the $i$-th component of an electric/magnetic dipole at  $\mathbf{r}_2$. This relation shows that a purely electric or magnetic source can be exchanged with its respective field at the observation point. The relationship is not as straightforward for an electric/magnetic field radiating from a magnetic/electric dipole. If we let $\mathcal{G}_{ij}^{\mathrm{HP}}$ define a Green's tensor encapsulating a magnetic field from an electric dipole, and $\mathcal{G}_{ij}^{\mathrm{EM}}$  contain an electric field from a magnetic dipole, the relation reads (see, e.g. \cite{Miller:EECS-2012-115}):}
\begin{align}
\highlight{\mathcal{G}_{ij}^{\mathrm{EM}}(\mathbf{r}_{2},\mathbf{r}_{1}) =- \mathcal{G}_{ji}^{\mathrm{HP}}(\mathbf{r}_{1},\mathbf{r}_{2}).\label{EH}}
\end{align}
Equations (\ref{EE}) and (\ref{EH}) enable us to simplify the expression describing the interplay between the fields and our goal structure. This is discussed in detail in the next section. 
\subsection{Choosing a merit function}
To create an algorithm capable of producing a helicity lattice with a desired geometrical pattern, we need to choose a merit function that will output a number that encodes the closeness of a given field to the goal, which we will aim to optimize. Inspired by \cite{MieScat}, wherein a method of controlling 3D optical fields and their intensity is proposed, we seek to control the value of helicity density at points in 3D space. Since we possess a goal pattern $\mathcal{H}_{0}$ \highlight{given by the helicity lattices reported in \cite{jorg}}, a natural choice would be to define a set $S$, of $N$ critical points, where $\mathcal{H}_{0}$ is ``close" to maximum and minimum values respectively. A simulated helicity density $\mathcal{H}[\mathbf{E},\mathbf{H}]$, can be compared with the values $\mathcal{H}_{0}$ takes at these coordinates; hence a measure of similarity between the two fields can be established. \highlight{Assuming that our goal structure lies in the plane $z=z_{0}$, we define the set of points $S$, as a collection of $N$ tuples of the form $\mathbf{s}_{i}=\{x_{i},y_{i},z_{0}\}$. We then seek points $\mathbf{s}_{i}$ for which $\mathcal{H}_{0}(\mathbf{s}_{i})$ is close to its minimum and maximum. We can then define an arbitrary bound $a \in (0,1]$ and form the set $S$ of size $N$, such that $\mathbf{s}_{i} \in S $ if
\begin{align}\label{bounds}
\mathcal{H}_{0}(\mathbf{s}_{i}) < a\cdot\mathrm{\mathrm{min}}[\mathcal{H}_{0} ] \text{ or } \mathcal{H}_{0}(\mathbf{s}_{i}) > a\cdot\mathrm{\mathrm{max}}[\mathcal{H}_{0} ].
\end{align}
Our simulation space is discretised, thus the $N$ is finite and will depend on the size of the domain and the  resolution, as well as the value of $a$.}
The merit function $\Theta[\mathbf{E},\mathbf{H}]$ will thus take the form:
\begin{align}
\Theta[\mathbf{E},\mathbf{H}] =\simpleSum \highlight{\theta_{k}^{2}(\mathbf{s}_{k}),}
\end{align}
where,
\begin{align}\label{theta_k_def}
\highlight{\theta_{k}(\mathbf{s}_k) \equiv \mathcal{H}[\mathbf{E},\mathbf{H}](\mathbf{s}_{k})- \mathcal{H}_{0}(\mathbf{s}_{k}),}
\end{align} 
is the error between the goal pattern and the simulated field at a coordinate $\mathbf{s}_{k}$. The closer $\Theta$ is to zero, the closer $\mathcal{H}[\mathbf{E},\mathbf{H}]$ will approximate the desired pattern $\mathcal{H}_{0}$. 
\highlight{The merit function $\Theta$ is a sum of $N$ functionals $\theta_{k}^{2}$}, that are evaluated at $\mathbf{s}_{k}$; by linearity, their variation can be evaluated as
\begin{align}
\delta \Theta[\mathbf{E},\mathbf{H}]& = \delta \simpleSum \theta_{k}^{2}(\mathbf{s}_k ) = 2 \simpleSum \theta_{k}(\mathbf{s}_k)\delta\theta_{k}(\mathbf{s}_k) \notag \\
& = 2 \simpleSum\theta_{k}(\mathbf{s}_k)\delta\mathcal{H}[\mathbf{E},\mathbf{H}](\mathbf{s}_{k}).
\end{align}
\highlight{We want to calculate the $\delta \mathcal{H}[\mathbf{E},\mathbf{H}]$ at each site in $S$.  $\mathcal{H}[\mathbf{E},\mathbf{H}]$ is dependent on the complex fields $\mathbf{E}$ and $\mathbf{H}$, suggesting that we write}
\begin{align}\label{delta_h_expand}
	\delta\mathcal{H}[\mathbf{E},\mathbf{H}] =\,& \pardev{\mathcal{H}}{\mathbf{E}}\cdot \delta\mathbf{E} + \pardev{\mathcal{H}}{\mathbf{E}^{*}}\cdot\delta\mathbf{E}^{*} \notag \\
&+ \pardev{\mathcal{H}}{\mathbf{H}}\cdot\delta\mathbf{H} + \pardev{\mathcal{H}}{\mathbf{H}^{*}}\cdot\delta\mathbf{H}^{*}. 
\end{align}
By plugging in the definition of helicity density from Eq.~\eqref{helicity} into Eq.~\eqref{delta_h_expand} and simplifying, we obtain\highlight{
\begin{align}
\delta \mathcal{H}[\mathbf{E},\mathbf{H}] =-\dfrac{i}{2 c\omega} \mathrm{Re}\left\{\mathbf{H^{*}}\!\cdot\delta \mathbf{E} - \mathbf{E^{*}}\!\cdot\delta\mathbf{H} \right\},
\end{align}
where the all fields are evaluated at $\mathbf{s}_{k}$.} 

\highlight{The final step in the method is calculating $\delta\mathbf{E}$ and $\delta\mathbf{H}$. We employ an approach called shape calculus, which is established on understanding the relationship between Maxwell's equations and small changes in geometry \cite{ShapeDO}. In our present problem, this manifests as the link between introducing small pieces of dielectric material and their effect on the existing electric and magnetic fields. The principle behind evaluating these can be illustrated with the example of the electric field $\mathbf{E}$. The effect of modifying the geometry by adding a piece of a polarizable dielectric will cause a change $\delta \mathbf{E} = \mathbf{E}_{\mathrm{after}}-\mathbf{E}_{\mathrm{before}}$ in the electric field, with a similar argument applying for the magnetic field $\mathbf{H}$; $\delta \mathbf{H} =  \mathbf{H}_{\mathrm{after}}-\mathbf{H}_{\mathrm{before}}$. As discussed in detail in, for example,  \cite{Miller:EECS-2012-115},} the variations in the fields $\mathbf{E}$ and $\mathbf{H}$ are given by a product of a Green's tensor $\mathcal{G}$ with a source current $\mathbf{P}$,
\highlight{
\begin{align}
\delta \mathbf{E}(\mathbf{s}_{k}) =& \int_{\phi} \mathrm{d}^{3} \mathbf{r} \ \mathcal{G}^{\mathrm{EP}}(\mathbf{s}_{k},\mathbf{r})\mathbf{P}(\mathbf{r}) \label{GT}\\ 
\delta \mathbf{H}(\mathbf{s}_{k})	 =& \int_{\phi} \mathrm{d}^{3} \mathbf{r} \  \mathcal{G}^{\mathrm{HP}}(\mathbf{s}_{k},\mathbf{r})\mathbf{P}(\mathbf{r}) \label{GT2},
\end{align}}
where $\phi$ is a 3D region containing the inclusion.
Taking into account the reciprocity relations in Eq.~\eqref{EH}, we can write the variation in $\theta_{k}^{2}$ at a point $\mathbf{s}_{k}$  as:
\begin{align}\label{delta_int}
\delta \theta_{k}^{2}(\mathbf{s}_{k}) =-\dfrac{i}{2 c\omega} 
\int_{\phi} \mathrm{d}^{3} \mathbf{r} \
\mathrm{Re}
\left\{
\mathbf{P}(\mathbf{r})\cdot \mathbf{\mathbf{F}}_{k}(\mathbf{r},\mathbf{s}_{k})
\right\},
\end{align}
where we defined;
\begin{align}\label{F}
\mathbf{\mathbf{F}}_{k}(\mathbf{r},\mathbf{r}') =  [ \mathbf{H}^{*}(\mathbf{s}_{k})&\mathcal{G}^{\mathrm{EP}}(\mathbf{r},\mathbf{r}')\notag \\
&+ \mathbf{E^{*}}(\mathbf{r}')\mathcal{G}^{\mathrm{EM}}(\mathbf{r},\mathbf{r}')]\theta_{k}(\mathbf{r}').
\end{align}
\highlight{Here lies the most important step of the adjoint method; since the structure of Eq.~(\ref{F}) is that of an electric field, that is, a Green's tensor multiplying a source current at the observation point $\mathbf{s}_{k}$, $\mathbf{F}_{k}(\mathbf{r},\mathbf{s}_{k})$ is the $k$th adjoint field. Exploiting the source-observer symmetry in reciprocal media, each source at $\mathbf{s}_{k}$ provides us with a field profile at $\mathbf{r}$. }

We can further simplify Eq.~(\ref{delta_int}) by \highlight{taking the dielectric inclusion's relative permittivity $\varepsilon_{\mathrm{I}}$ to be homogenous and isotropic}. The induced polarizability, $\mathbf{P}(\mathbf{r})$ can then be written as
\highlight{\begin{align}
\mathbf{P}(\mathbf{r})= \Delta \varepsilon \mathbf{E}(\mathbf{r}),
\end{align}
where $\Delta\varepsilon \equiv \varepsilon_{\mathrm{I}}- \varepsilon^{(0)}$, is the difference in relative permittivity between the inclusion and the background $\varepsilon^{(0)}$. Moreover, we assume that the inclusion is sufficiently small that we can take its centre to be at $\mathbf{r'}$, thus approximating the fields in the integrand of Eq.~(\ref{delta_int})  by their value at $\mathbf{r'}$. This gives}
\begin{align}
\delta \theta_{k}^{2}(\mathbf{s}_{k}) \approx -\dfrac{i \Delta\varepsilon V}{2 c\omega} 
\mathrm{Re}
\left\{
\mathbf{E}(\mathbf{r'})\cdot \mathbf{\mathbf{F}}_{k}(\mathbf{r'},\mathbf{s}_{k})
\right\},
\end{align}
where $V$ is the volume of the inclusion.
 \highlight{We are only interested in relative changes in the merit function between sites, thus the scalar prefactor $\left( -i\Delta\varepsilon V/2c\omega\right)$ can be ignored.}  This allows us to finally write the variation in $\theta^{2}_{k}$ at a site $\mathbf{s}_{k}$ as
\begin{equation}\label{dthetaj}
\delta \theta^{2}_{k}(\mathbf{s}_{k})  \approx  
\mathrm{Re}
\left\{
\mathbf{E}(\mathbf{r'})\cdot \mathbf{\mathbf{F}}_{k}(\mathbf{r'},\mathbf{s}_{k})
\right\},
\end{equation}
and by linearly adding $N$ adjoint fields we have for the total variation:
\begin{align}
	\delta \Theta\left[ \mathbf{E},\mathbf{H}\right]  & \approx
	\mathrm{Re}
	\left\{
	\mathbf{E}(\mathbf{r'})\cdot \sum_{k=1}^{N}  \mathbf{\mathbf{F}}_{k} (\mathbf{r'},\mathbf{s}_{k})
	\right\}\notag \\ 
	& = \label{EA1} \mathrm{Re}\left\{ \mathbf{E}(\mathbf{r'})\cdot \mathbf{F}(\mathbf{r'})\right\}.
\end{align}
The change (variation) in the merit function $\Theta$ is thus given by the overlap between the \highlight{forward} field $\mathbf{E}$ and the adjoint field $\mathbf{F}$. 

\subsection{Advantages of adjoint method over brute force optimization}
\begin{figure}[h]
	\centering
	\includegraphics[width=1\linewidth]{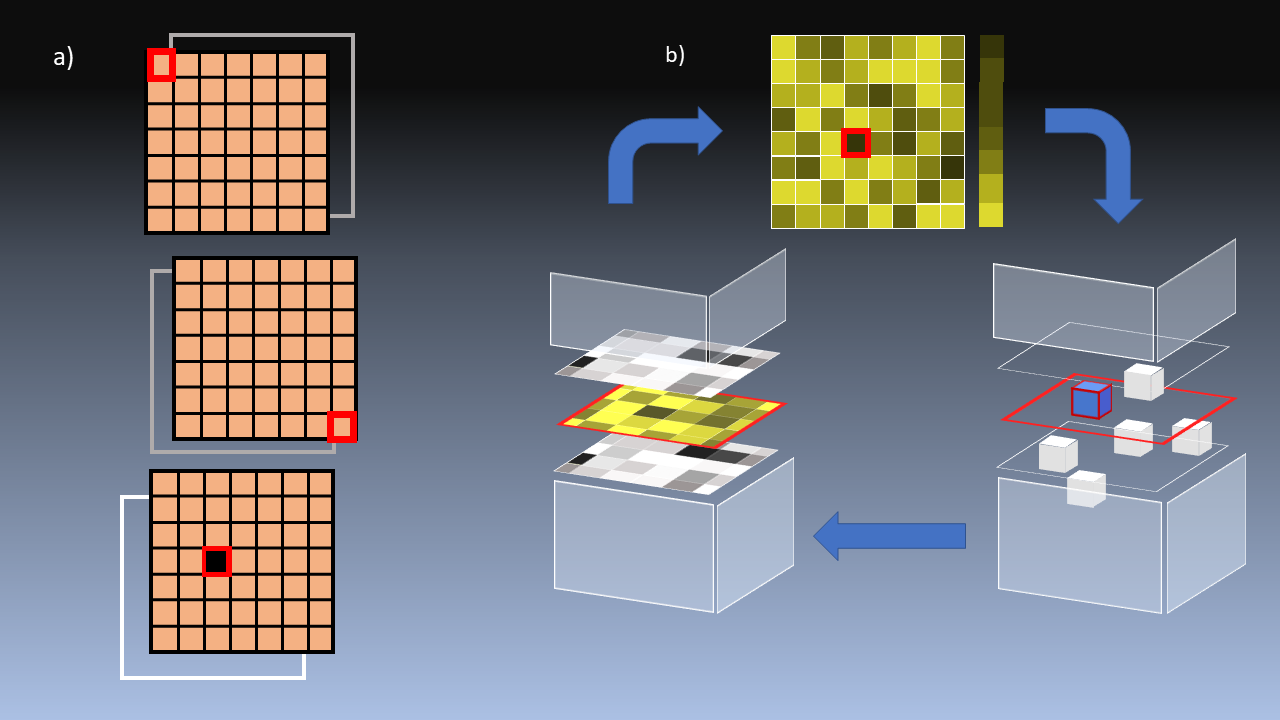}
	\caption{A brute-force optimization in a) and adjoint method in b). In a) the algorithm has to iterate through every coordinate in 3D space placing a test inclusion there and running a separate simulation before it can determine the optimal location. In b) the adjoint method finds this coordinate by doing only two simulations. Both a) and b) continue until the merit function value can no longer decrease.}
	\label{fig:iditerationprocedure}
\end{figure}
To find a structure producing a helicity lattice, one might either use intuition or follow a brute-force process of building it. The latter involves trying every coordinate one by one and measuring what effects placing the inclusion there has on the merit function. One then would narrow it down to the single coordinate where the dielectric inclusion exerts the largest influence and place it there. The whole procedure would have to be repeated for every element added to the structure until the goal field is reached. A 3D simulation would require a number of iterations equal to the number of inclusions needed to reach the goal, multiplied by $n^{3}$, where $n$ is the number of grid points along the side of the simulation box.

By employing the adjoint method instead, we no longer need to check each coordinate to see how the function responds to adding an inclusion. This information is contained in the $\delta \Theta$, an array of values corresponding to each coordinate in the domain. By picking the highest value of the array, i.e. placing the dielectric at the coordinate with the greatest number, we are guaranteed to produce the highest change in the function $\Theta$. This process saves computational resources and makes certain optimization goals feasible by \highlight{dramatically decreasing simulation time}. 

\highlight{The comparison between the brute force and adjoint approaches is illustrated schematically in Fig.~\ref{fig:iditerationprocedure}. The brute force approach (left) scans each coordinate and measures the impact on the merit function by placing a test inclusion there. Only after finding the location with the highest influence the algorithm commits to placing the inclusion permanently.} The adjoint method (right) scans the 3D space for the highest value of $\delta \Theta$ from Eq.~(\ref{EA1}). It then places an inclusion there, circumventing the need for additional simulations.

\section{Procedure}\label{ProcedureSection}
\begin{figure*}
	\centering
	\includegraphics[width=1\linewidth]{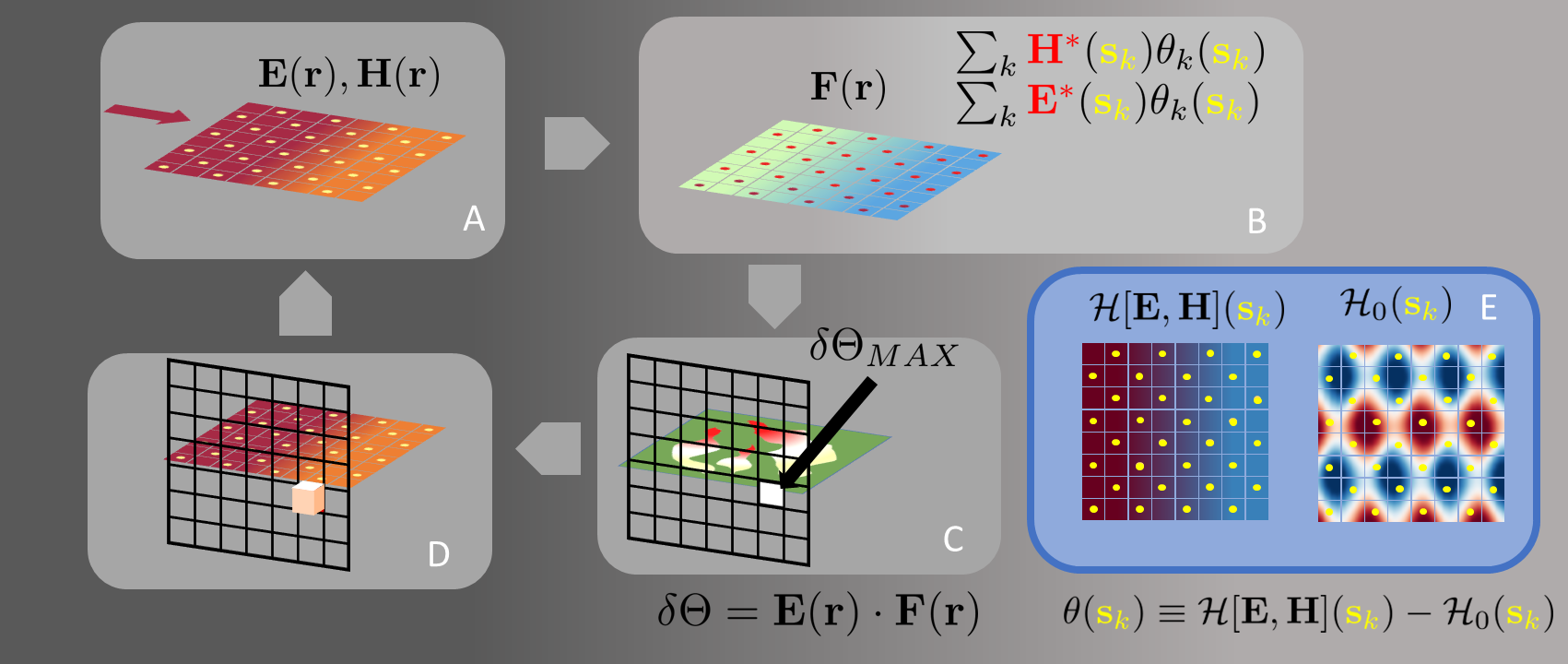}
	\caption{\highlight{Schematic illustration showing the adjoint optimisation cycle (boxes ABCD), and the depiction of the comparison between the goal structure and the simulated field in E. The control points are depicted in yellow, both pictorially, and symbolically, i.e., $\mathbf{s}_{k}$, inside the expressions. The red dots, along with the red colouring of the fields' $\mathbf{E}^{*}$ and $\mathbf{H^{*}}$symbols, signify the sources of the adjoint field (situated at the yellow control points). Starting the cycle at A, we obtain the values of the fields $\mathbf{E}(\mathbf{s}_{k})$ and $\mathbf{H}(\mathbf{s}_{k})$ at the yellow control points, arising due to the single light source. In B, we use the field values $\mathbf{E}(\mathbf{s}_{k})$ and  $\mathbf{H}(\mathbf{s}_{k})$ from A, and use them as amplitudes of the adjoint dipoles (in red) that we excite at every control point $\mathbf{s}_{k}$, to produce the adjoint field $\mathbf{F}$. Then in C, we combine the fields from A and B obtaining the derivative of the merit function, i.e., $\delta\Theta=\mathbf{E}\cdot\mathbf{F}$, and find the coordinate where the value of $\delta \Theta$ is the highest. In D, we place the dielectric inclusion at this coordinate, and re-run the simulation in A with the updated geometry. The cycle repeats until the value of the merit function $\Theta$ is sufficiently low, i.e., the fields in the box E look alike. The mesh in C and D shows that a maximum value of $\delta \Theta$ is sought across the 3D space; the algorithm is not favouring any particular plane, only a south facing wall is being shown for clarity of the image.} } 
	\label{fig:idstructure}
\end{figure*}

We aim to create a two-dimensional pattern embedded in a three-dimensional region; it is, therefore, sufficient to generate EM waves that agree with the goal lattice only at the particular plane where the structure resides (say, $z=0$). Every point in space apart from the region immediately surrounding the observation slice is thus a potential target for the placement of a dielectric inclusion. Since we want our structure to allow for physical access to the helicity lattice, we leave a margin of width $l$ at each side of the $z=0$ slice. 
The simulation space is discretized in units of wavelength; any design that we produce can be adapted to an arbitrary length regime as Maxwell's equations are scale invariant \highlight{(in the absence of charges or currents, as is the case here)}.
\highlight{We define our optimisation target as $\mathcal{H}_{0}$, which is the normalized (between $-1$ and $1$), analytical expression for the chosen helicity lattice. We will choose the parameter $a$, in line with Eq.~(\ref{bounds}), effectively controlling the tolerance in any small mismatch between the lattice produced and the goal.} 

The first step of the optimization process is to run the simulation in the specified region and record the values of the steady-state fields $\mathbf{E}$ and $\mathbf{H}$, \highlight{ as shown in the box A in Fig.~\ref{fig:idstructure}}. We then construct the sources for the adjoint simulation \highlight{(box B in Fig.~\ref{fig:idstructure})}; at each site $\mathbf{s}_{k}$, we place dipoles whose amplitude depends on the steady-state values of the fields and the pattern defined at $\mathbf{s}_{k}$, i.e. $\mathbf{E}(\mathbf{s}_{k}),\mathbf{H}(\mathbf{s}_{k})$ and $\mathcal{H}_{0}(\mathbf{s}_{k})$. At each $\mathbf{s}_{k}$, we position both an electric dipole with the amplitude $\theta_{k}(\mathbf{s}_{k}) \mathbf{H^{*}}(\mathbf{s}_{k})$, and a magnetic dipole with the amplitude $\theta_{k}(\mathbf{s}_{k}) \mathbf{E^{*}}(\mathbf{s}_{k})$, where $\theta(\mathbf{s}_{k})$ \highlight{is defined in Eq.~\eqref{theta_k_def} and depicted schematically in box E of Fig.~\ref{fig:idstructure}}. Running the backward (adjoint) simulation concludes the first step. Upon recording the values of the steady-state fields $\mathbf{F}$, we combine them with the values of $\mathbf{E}$ via Eq.~(\ref{EA1}). The result encapsulates the relationship between the placement of a dielectric inclusion and a corresponding change to the merit function $\Theta$. In other words, picking a coordinate $\mathbf{r}_{\mathrm{max}}$ such that $\mathrm{Re}\left\{ \mathbf{E}(\mathbf{r}_{\mathrm{max}})\cdot \mathbf{F}(\mathbf{r}_{\mathrm{max}})\right\}$ is maximum, ensures that placing an inclusion at $\mathbf{r}_{\mathrm{max}}$ will result in the largest change in $\Theta$ \highlight{(boxes C and D in Fig.~\ref{fig:idstructure})}. We then update the geometry and repeat the previous step until the desired pattern has been achieved.

\highlight{The overall algorithm} can be summarized in the following way;
\begin{enumerate}
	\item First stage:
	\begin{enumerate}[label=(\roman*)]
		\item Define a 3D simulation space; choose a plane where the 2D helicity pattern is to be created and situate its normalized analytical expression there.
		\item In the chosen plane, pick a set $S$ of critical points $\mathbf{s}_{k}$, where the analytical function is ``close" to its maximum absolute values \highlight{(box E in Fig.~\ref{fig:idstructure})}. 
		\item By running the forward simulation, obtain steady-state values of the fields $\mathbf{E}$ and $\mathbf{H}$ at every point in the 3D domain, including the points of interest $\mathbf{s}_{k}$ \highlight{(box A in Fig.~\ref{fig:idstructure})}.
		\item Place electric and magnetic dipoles at every $\mathbf{s}_{p}$ as functions of the $\mathbf{E}(\mathbf{s}_{p})$ and $\mathbf{H}(\mathbf{s}_{p})$ at the first step. Obtain the steady state values of the fields by running the adjoint simulation \highlight{(box B in Fig.~\ref{fig:idstructure})}.
		\item  Combine the forward and adjoint fields, as in Eq.~(\ref{EA1}). Pick the coordinate where the $\delta \Theta$ is the highest and place the dielectric inclusion there \highlight{(boxes C and D in Fig.~\ref{fig:idstructure})}.
	\end{enumerate}
	\item Second stage (iterative):
	\begin{enumerate}[label=(\roman*)]
		\item Repeat the steps (iii)-(v) from the first stage.
		\item Repeat until the forward field approximates the desired pattern at the chosen plane.
	\end{enumerate}
\end{enumerate}
\section{Results}\label{ResultsSection}
Using the single beam technique supported by the adjoint method, we have created the patterns approximating the helicity density originally arising as a superposition of three waves in two different configurations. \highlight{While we have chosen just two}, many other non-interfering superpositions are realizable, and the list of their explicit constituent light sources can be found in \cite{jorg}. 

\highlight{The parameters defining the simulations are as follows. All distances in all computations are expressed in units of wavelength, and each has a chosen resolution parameter $R$ such that there are $R$ pixels per wavelength. The dielectric inclusions are chosen to be cubic with side length equal to $1/R$ wavelengths (i.e., one pixel) and are referred to hereafter as `blocks'. In all simulations the relative permittivity of the blocks was $\varepsilon = 1.3$, the coefficient $a$ in Eq.~\eqref{bounds} was set to $0.3$, the gap $l$ was chosen to be three blocks (i.e. three pixels, or $3/R$ wavelengths), and the incident plane wave is $y$-polarised and propagating in the $x$ direction. All simulations were performed using the FDTD library MEEP \cite{Meep}, within which we used a built-in implementation of perfectly matched layers \cite{berengerPerfectlyMatchedLayer1994} as boundary conditions at the edge of the simulation box. The underlying code, along with  detailed documentation can be found online at Ref.~\cite{kilianski2022inversedesign}.}
\subsection{Rectangular three wave superposition} 
\begin{figure}[h]
	\centering
	\includegraphics[width=1\linewidth]{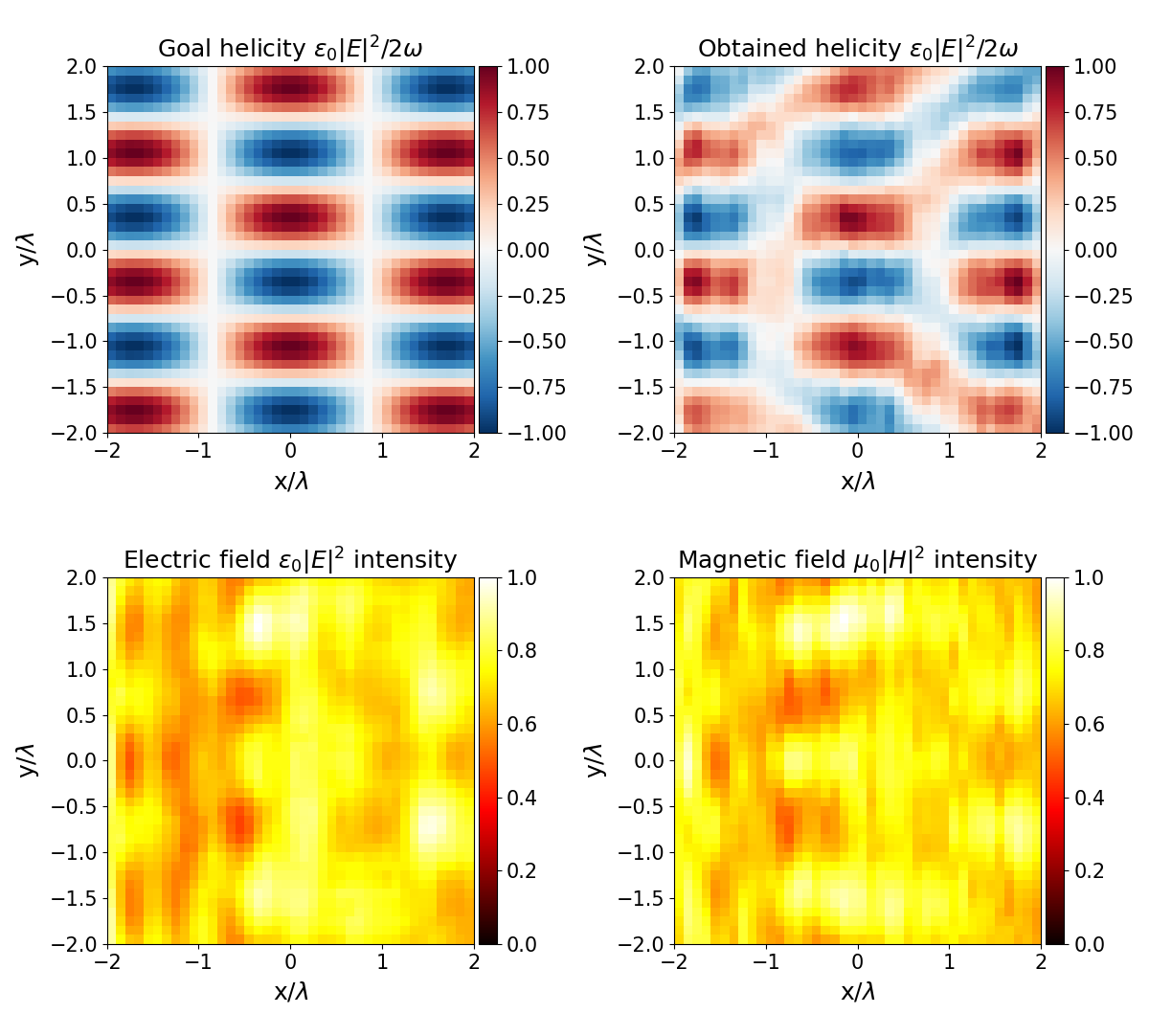}
	\caption{Comparison between the goal pattern (upper left panel) and that obtained by using an inverse-designed structure and a single source (upper right panel) for an originally three-source helicity lattice \highlight{in a checkerboard configuration}. The lower panels show the obtained field's \highlight{normalised} inhomogeneity in the square of the electric (left) and magnetic (right) fields, respectively. The simulation was run for 4000 iterations with resolution of 10 pixels per wavelength $\lambda$.}
	\label{fig:sc3200res6}
\end{figure}

\highlight{We first use the above-described method to produce the simplest two-dimensional helicity lattice, namely a rectangular three-wave superposition that results in a checkerboard pattern (see Table.~2 and Fig.~3 in \cite{jorg}), but here from a single plane wave input. The goal and optimised patterns are shown in the upper two panels in Fig.~\ref{fig:sc3200res6}, with the lower two panels here (and in the two subsequent figures) included for later discussion in Section \ref{analysisSection}}. 
We observe that since the source plane wave oscillates perpendicular to the lattice's vertical lines of symmetry, the resulting simulation approximates the helicity lattice well (showing clear periodicity by agreeing with the goal pattern at the $\mathcal{H} = 0$ points) even in lower resolution settings (i.e.~less than six pixels per wavelength). 
\highlight{The reason for this is suggested by noting that} the goal structure $\mathcal{H}_{0}$, has the form:
\begin{align}
\mathcal{H}_{0} = \mathcal{N}\cos\big[(2 - \sqrt{2}) \pi x\big]\sin(\sqrt{2} \pi y),\label{h1}
\end{align}
where the factor $\mathcal{N}$ carries the units of $\varepsilon_{0}|E|^{2}/ 2\omega$. This lattice has two orthogonal lines of symmetry: $x = (2-\sqrt{2})n/2$, and $y = n/\sqrt{2}$, $n \in \mathbb{N}$, where $\mathcal{H}_{0}$ is zero.  The algorithm can thus independently influence the fields along those degrees of freedom more ``naturally", resulting in the outline of the correct structure emerging at a very early point of the simulation ($< 100$ iterations, \highlight{well before the 4000th iteration shown in Fig.~\ref{fig:sc3200res6}}).

\subsection{Triangular (diamond) three-wave superposition}
\begin{figure}[h]
	\centering
	\includegraphics[width=1\linewidth]{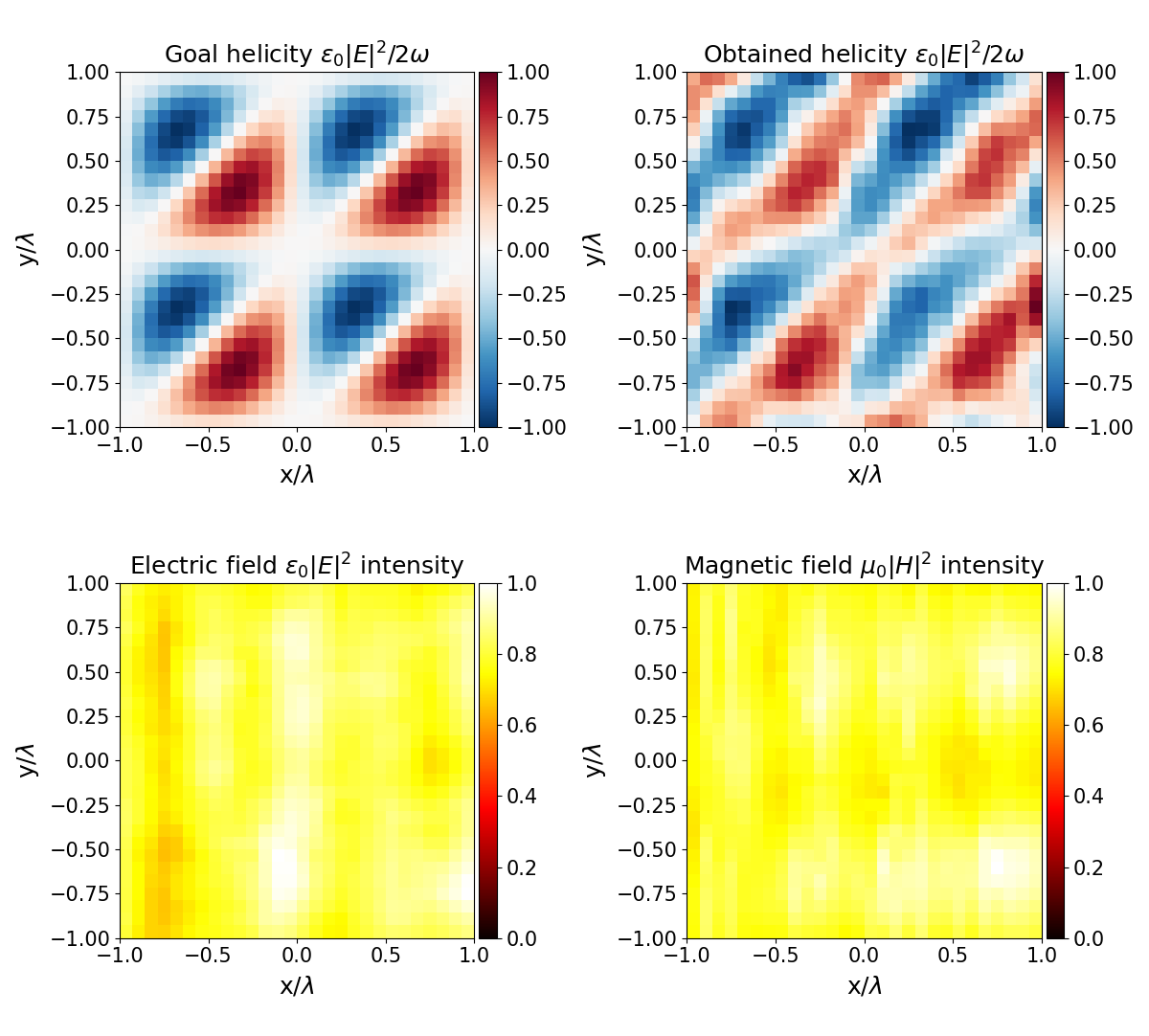}
	\caption{Comparison between the goal pattern (upper left panel) and that obtained by using an inverse-designed structure (see Fig.~\ref{fig:compare3d}a) and a single source (upper right panel) for an originally three-wave helicity lattice \highlight{in a triangular (diamond) configuration}. The lower panels show the obtained field's \highlight{normalised} inhomogeneity in the square of the electric (left) and magnetic (right) fields, respectively. The simulation was run for 2000 iterations with a resolution of 14 pixels per wavelength $\lambda$.}
	\label{fig:vec3142000}
\end{figure}

The same number of sources as in the previous example, albeit angled differently \cite{jorg} produces a new lattice. The analytical expression in this case reads:
\begin{align}
\mathcal{H}_{0} = \mathcal{N} \left\{-\sin(2 \pi x) + \sin[2 \pi (x-y)]  + \sin(2 \pi y)\right\}\label{h2},
\end{align}
The results of the optimisation are plotted in Fig.~\ref{fig:vec3142000},
and for this calculation we also show in Fig.~\ref{fig:compare3d}a the geometry of the resulting 3D dielectric crystal \highlight{which, as anticipated, shows no recognisable structures that could have been reached by intuition}.

\begin{figure}[h]
	\centering
	\includegraphics[width=1\linewidth]{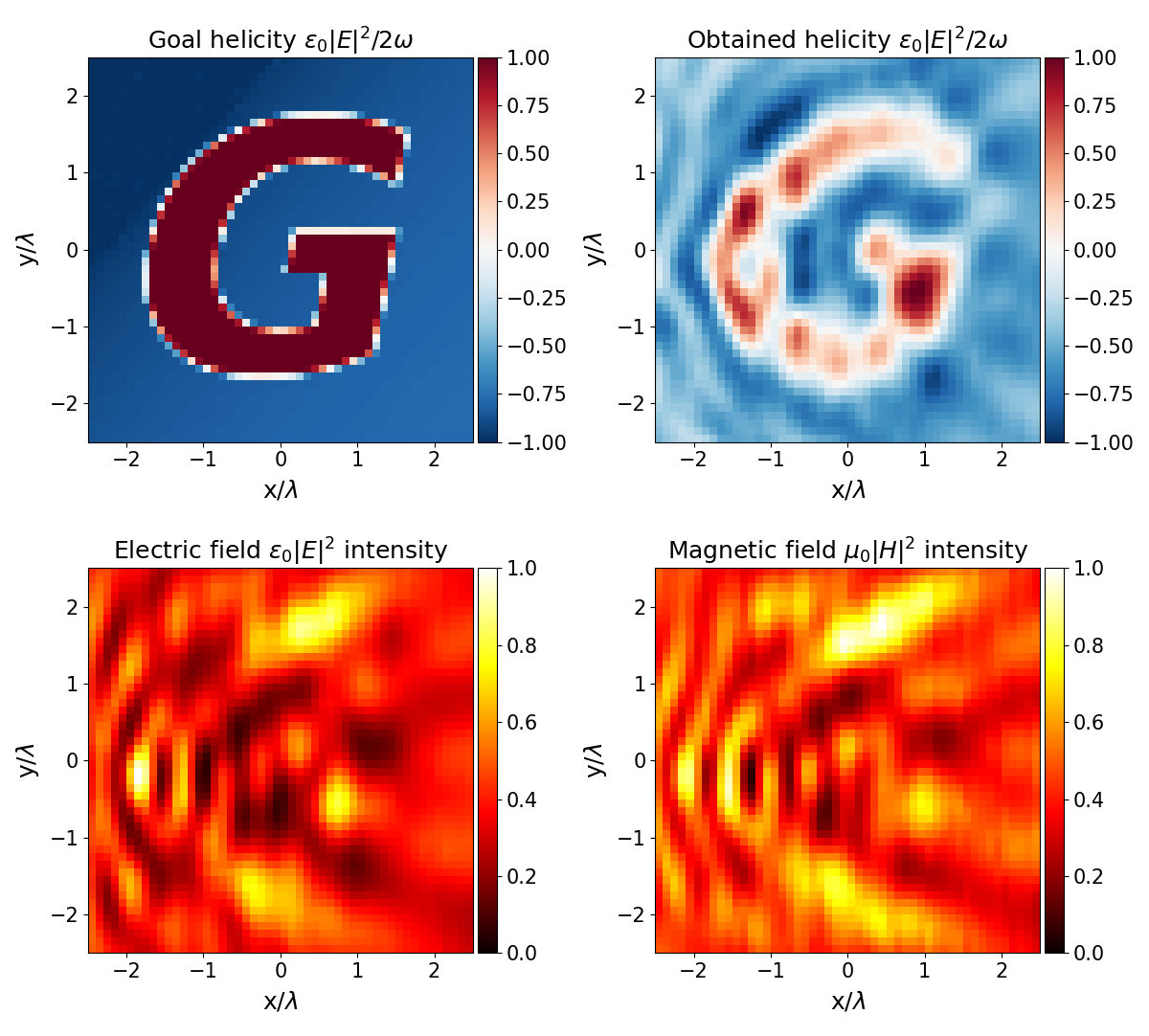}
	\caption{Comparison between an arbitrary user-defined helicity image (upper left panel) and that obtained by using an inverse-designed structure (see Fig.~\ref{fig:compare3d}b) and a single source (upper right panel) producing a helicity density pattern in the shape of a letter ``G" . The lower panels show the obtained field's \highlight{normalised} inhomogeneity in the square of the electric (left) and magnetic (right) fields, respectively. The simulation was run for 16000 iterations with a resolution of 10 pixels per wavelength $\lambda$.}
	\label{fig:arbitraryG}
\end{figure}
\begin{figure*}
	\centering
	\includegraphics[width=0.9\linewidth]{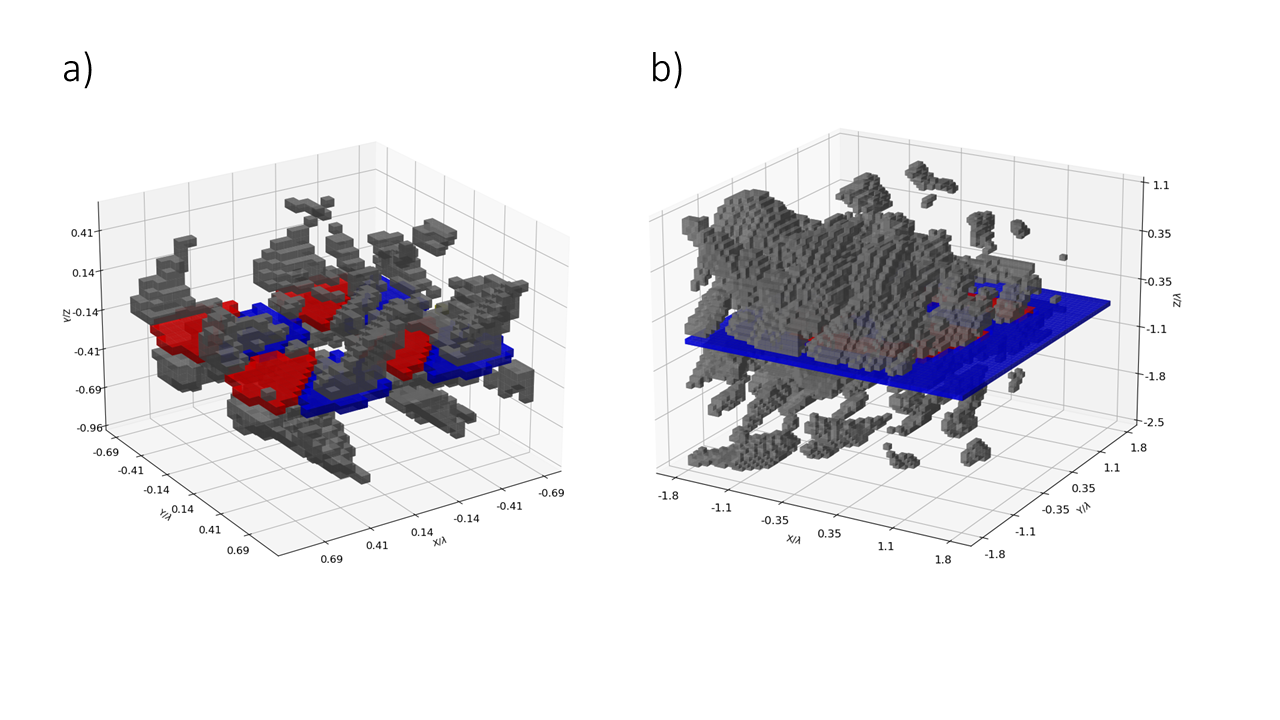}
	\caption{The three-dimensional models of the crystal-like structure. In a), the arrangement produces the three-wave triangular (diamond) lattice \highlight{as shown in Fig.~\ref{fig:vec3142000}}. In b), the structure produces an arbitrary pattern (letter ``G") \highlight{as shown in Fig.~\ref{fig:arbitraryG}}. In both cases the gap at the level $z=0$ is three blocks high.}
	\label{fig:compare3d}
\end{figure*}

In contrast to that described by Eq.~(\ref{h1}), this lattice possesses more than two lines of symmetry. This can be inferred from inspecting the arguments of the $\sin$ function in (\ref{h2}); for $n \in \mathbb{N}$, lines $x = n/2$,$y = n/2$ and $y = x - n/2$ are the zeros of $\mathcal{H}_{0}$. As a result, it takes more iterations for the pattern to become apparent as the initial blocks force the simulated lattice to have the zeros at $y = x - n/2$, exploiting the wave nature of the source and modulating the helicity density in the diagonal direction. The other lines of symmetry emerge later as more blocks are added to the structure, approximating the original lattice to a precision controlled by the overall resolution (number of pixels per wavelength). 

\subsection{Arbitrary helicity pattern}
\highlight{Although our technique has shown itself to be effective in constructing helicity lattices, there is nothing in principle tying it to a regular, repeating pattern. The algorithm understands any helicity pattern as a series of pixels, which can be viewed as an input to the algorithm and therefore chosen arbitrarily. To demonstrate this, we constructed a greyscale image of the letter ``G", whose grey values were then normalised to lie between $-1$ and $1$  \highlight{in the same way as the lattices previously investigated}. The algorithm was tasked with reproducing this image; the results are shown in Fig.~\ref{fig:arbitraryG} and the corresponding structure is shown in Fig.~\ref{fig:compare3d}b}. We note that the algorithm correctly distributes the regions of positive and negative helicity density (the inside of the G is positive), whereas the outline of the letter has helicity density close to zero. The shape itself is recognisable; however, some distortion occurs and artifacts are present \highlight{even though the simulation was left to run for much longer than the lattices discussed in the previous two sections (16000 iterations, rather than 4000 and 2000, respectively)}. The process constructs the base and the top of the G first \highlight{as these align with oscillations of the source field}; the ``ripples" of positive helicity spread to the bounds of the shape, and further details are added successively later. The size of the simulation volume constrains the quality of the resulting lattice; such images without periodicity and increased level of detail require a large computational volume. Additionally, if one was mostly concerned with the fidelity of the reproduced image, another input beam could be incorporated into the process to improve the quality and speed up the process.

\section{Analysis}\label{analysisSection}
In the case of analytical solutions for helicity lattices (four sources or fewer), the electric field intensity can always be made homogeneous \cite{jorg}. This is not always the case for simulated structures, \highlight{as can be observed in Figs.~\ref{fig:sc3200res6}-\ref{fig:arbitraryG}}. This important detail influences their potential experimental use. \highlight{As mentioned in the introduction, forces on} atoms and molecules in trapping potentials is constrained by optical forces induced by gradients in the electric intensity. Thus, to create an efficient helicity lattice, one would require the absence of intensity gradients in the region spanned by the structure to allow for interrupted chiral interactions between the enantiomers and the lattice. The fields generated by our simulation possess inhomogeneities due to the finite width of the plane waves generating the helicity. \highlight{For the lattices shown in Figs.~\ref{fig:sc3200res6} and \ref{fig:vec3142000},} the inhomogeneity is dependent on the size of a computational vs observational region; the larger the gap between the boundaries of a computational volume and an observation box, the closer the source resembles a plane wave. This, in turn, produces an electric field intensity with smaller variations; an infinitely sized computational box would allow one to obtain a perfectly homogenous intensity. Consequently, a trade-off exists between a reasonable computational time and fidelity of the lattice compared to the goal structure. Routes towards reducing \highlight{the larger} inhomogeneity in intensity exhibited by arbitrary patterns \highlight{such as that shown in Fig.~\ref{fig:arbitraryG} are not clear to us}. Since they are artificially introduced as a result of interacting fields, the relationship between their shapes and the resulting intensity gradients is unclear. Future work will address this issue and focus on developing more sophisticated methods for producing an arbitrary pattern. An additional merit function could be defined, specifically controlling the intensity gradients. Additionally, a weight function between the two merit functions would be introduced. A link could potentially be established between certain shapes and the homogeneity of the intensity; this could be revealed by a machine-learning algorithm. 
 
As can be seen in Fig.~\ref{fig:compare3d}, some of the elements of the crystal are ``floating". \highlight{This is a particularly extreme example of the structural integrity issues that arise in general in inverse design (see, e.g.~\cite{AugenRock})}. Such problems can be rectified by running a longer simulation, effectively allowing the algorithm to fill the larger gaps by chance and then manually removing any remaining ``islands" of the dielectric material. Alternatively, a weight function can be implemented to penalise the algorithm for placing isolated blocks. A similar issue occurs as the structure comprises two halves, allowing access to the lattice; the crystal can be grown within the bounds of an existing structure providing support, or additional dielectric elements can be incorporated to join the two halves near the edges. \highlight{When considering manufacturing challenges for structures such as this, it is worth recalling that Maxwell's equations (without charges or currents) are scale-invariant, and therefore so are the helicity lattices and the structures proposed here. Thus, at this proof-of-principle stage, the choice of the size regime will dictate the difficulty of potential manufacturing challenges.}
\\
\section{Summary and conclusions}\label{conclusionsSection}
 In this work, a gradient-based formalism was used to develop a proof-of-principle scheme capable of reproducing a helicity lattice pattern, only with a reduced number of input sources. The problem was cast as a minimum-seeking algorithm, iteratively growing a crystal-like structure. This object refracts a single beam in such a way as to reproduce the effect of superposing multiple beams, resulting in the formation of a helicity lattice. We have presented the results of reproducing two originally three-wave helicity lattices using a single source input. Additionally, we introduced a modified scheme where the pattern to be reproduced was not a result of superposing light sources but rather an arbitrary image provided by the user. This approach opens up a possibility for designing bespoke helicity patterns suited to the needs of a particular experiment; it is worth mentioning that using the same principles, a different merit function can be provided, \highlight{thus it appears to be possible to use versions of this method to achieve structures arising from various mutual arrangements of the electric and magnetic fields, not just those where a helicity structure arises.}

The simulations reproducing known helicity patterns serve as proof-of-principle concepts and demonstrate the feasibility of using a single light source and a refracting object to mimic the behaviour of a field composed of multiple sources. We have described how increasing the computational volume approximates the plane wave more closely and, as a result, produces an increasingly homogeneous intensity in the case of known helicity patterns. We outlined possible strategies to resolve the issue of intensity gradients in arbitrary patterns \highlight{bringing practical helicity lattices a step closer to reality}.
\begin{acknowledgments}
The authors acknowledge financial support from the UK Research and Innovation Council through a doctoral training programme grant (EPSRC/DTP 2020/21/EP/T517896/1), New Horizons grant  EP/V048449/1 and New Investigator grant EP/W016486/1. The authors also acknowledge fruitful discussions with Stephen M.~Barnett, Jörg B.~Götte, Ben W.~Butler, Thomas M.~Jones and Zhujun Ye. We also want to thank Claire Cisowski for testing the early version of the code, Owen Miller for enlightening correspondence and Neel MacKinnon for critical reading of the manuscript. \\
\end{acknowledgments}

\end{document}